\documentclass[aps,prc,preprint,groupedaddress,showpacs]{revtex4-1}
\usepackage{graphicx}
\usepackage{amsmath}
\usepackage{color}
\allowdisplaybreaks

\begin{document}

\title{Semi-inclusive charged-current neutrino-nucleus reactions}

\author{O. Moreno}
\author{T. W. Donnelly}
\affiliation{Center for Theoretical Physics, Laboratory for Nuclear Science and Department of Physics, Massachusetts Institute of Technology, Cambridge, MA 02139, USA}
\author{J. W. Van Orden}
\affiliation{Department of Physics, Old Dominion University, Norfolk, VA 23529\\ Jefferson Lab, Newport News, VA 23606, USA}
\author{W. P. Ford}
\affiliation{Department of Physics, University of Southern Mississippi, Hattiesburg, MS 39406}

\date{\today}

\begin{abstract}
The general, universal formalism for semi-inclusive charged-current (anti)neutrino-nucleus reactions is given for studies of any hadronic system, namely, either nuclei or the nucleon itself. The detailed developments are presented with the former in mind and are further specialized to cases where the final-state charged lepton and an ejected nucleon are presumed to be detected. General kinematics for such processes are summarized and then explicit expressions are developed for the leptonic and hadronic tensors involved and for the corresponding responses according to the usual charge, longitudinal and transverse projections, keeping finite the masses of all particles involved. In the case of the hadronic responses, general symmetry principles are invoked to determine which contributions can occur. Finally, the general leptonic-hadronic tensor contraction is given as well as the cross section for the process.
\end{abstract}


\maketitle

% INTRODUCTION

\section{Introduction \label{introduction}}

Many of the ongoing experiments in charge-changing neutrino
scattering involve quasielastic scattering from light to medium mass
nuclei. An increasing number of these experiments offer the
possibility of studying semi-inclusive charge-changing (CC) neutrino
or antineutrino reactions, namely those where a final-state charged
lepton and some other particle are presumed to be detected in
coincidence.  For example, in the ArgoNeuT \cite{argoneut} and MicroBooNE
\cite{microboone} experiments protons together with muons
are detected in coincidence using argon TPCs. Using standard nuclear
physics notation such reactions would be denoted
$X(\nu_{\ell},\ell^-x)$ and $X(\bar{\nu}_{\ell},\ell^+x)$, where
$\ell=e$, $\mu$, or $\tau$. Here $x$ can be any kinematically
allowed particle, for instance, $\gamma$, a nucleon $N=p$ or $n$, a
deuteron $d$ or triton $t$, $^{3}$He, $\alpha$, fission fragment,
$\pi$, $K$, and so on. The target $X$ may be a nucleus or the proton
itself. All of these possibilities are contained in the formalism to
follow. One should be clear that this notation indicates what is
presumed to be detected, not what is actually in the final state.
For example, if $x=p$, this means that for sure one proton is in the
final state; however, depending on the kinematics chosen for the
reaction, there may be many open channels, a proton and a daughter
nucleus in some discrete state, two protons and a different nucleus
in some discrete state, a proton and a neutron and yet another
nucleus in some discrete state, {\it etc.} The semi-inclusive cross
section is then the sum/integral over all unobserved particles,
excepting only the one that is presumed to be detected, in this
example a proton. At a level lower, one has the inclusive cross section
where all particles for all open channels are to be
summed/integrated.

In the rest of the paper, to make things more specific and to
explore the case of most present interest in the quasielastic regime
(CCQE), we focus on the specific case of a nuclear target where a
nucleon is the particle that is presumed to be detected ($x=N$).
Nevertheless it should be clear that simply by changing the names of
the particles involved all of the developments can immediately be
used in any other semi-inclusive study. Accordingly we now consider
reactions of the type $^{A}_{Z}X(\nu_{\ell},\ell^-p)^{A-1}_{\quad
Z}Y$, $^{A}_{Z}X(\bar{\nu}_{\ell},\ell^+n)^{A-1}_{Z-1}Y$,
$^{A}_{Z}X(\nu_{\ell},\ell^-n)^{A-1}_{Z+1}Y$ and
$^{A}_{Z}X(\bar{\nu}_{\ell},\ell^+p)^{A-1}_{Z-2}Y$. These are to be
viewed in context with semi-inclusive electron scattering reactions
$^{A}_{Z}X(e,e^{\prime}p)^{A-1}_{Z-1}Y$ and
$^{A}_{Z}X(e,e^{\prime}n)^{A-1}_{\quad Z}Y$. In the initial state
one has some nucleus $X$ in its ground state with mass number $A$
and charge $Z$, while in the final state one has a nuclear system
$Y$ with mass number $A-1$ and the charges indicated above. The
latter daughter nucleus is not presumed to be in its ground state in
general (although this is one possibility when the system is stable
to nucleon emission) and may be in some discrete excited state (if
any exist), may be a granddaughter nucleus plus two nucleons, and so
on. All open channels are to be considered and we only require that
the mass number and charge be as indicated, together with the
kinematical information to be discussed in the following section.
Note also that of the four neutrino and antineutrino reactions given
above, the first two are in some sense ``natural'' in that the
reactions in the CCQE regime are at least dominated by the basic
reactions on nucleons in the target nucleus, namely, $\nu_{\ell}+n
\rightarrow\ell^-+p$ and $\bar{\nu}_{\ell}+p \rightarrow \ell^+ +n$,
respectively. However, the third and fourth reactions can occur in
nuclei. On the one hand, the final states involved are complex
interacting many-body states, involving in general coupled channels
whenever kinematically allowed. There may be several nucleons in the
final state and it is possible that one with the ``wrong'' flavor is
the one detected. In fact, for some situations there may be no bound
state of the final nucleus reached and one for sure has nucleons of
both flavors in the final state. On the other hand, while one
certainly has one-body electroweak current operators (those that act
on a single nucleon), it is also clear that two-body meson exchange
currents (MEC) are also present. For instance, an important
contribution to MEC at quasielastic kinematics are diagrams where
two nucleons interact with an exchanged $W^{\pm}$, going through a
virtual $\Delta$ which in turn exchanges a pion between the two
nucleons, leaving two nucleons in the final state. Take for
example the third reaction above: if the two initial nucleons are an
$nn$ pair in the nuclear ground state, one can absorb the exchanged
$W^+$, go through a $\Delta^+$, exchange a $\pi^+$, and have an $np$
pair in the final state where the neutron is the particle detected
in the third reaction (and the proton may be the one detected in the
first reaction). In the developments presented in the rest of this
paper the formalism is general enough to allow for MEC, no
assumption is required about which specific reaction is being
considered and only when applying these ideas with particular
modeling are the details required. All of the developments are kept
relativistic, {\it i.e.,} no non-relativistic approximations are
made, with one exception which will be discussed later in this
paper. All of the formalism may then be used regardless of the
energy scale, whether at relatively low energies or, what is more
typical, at high energies.

The paper is arranged in the following way: in the next section the required basic kinematics are summarized. Here we assume that the incident neutrino or antineutrino has a given momentum, although in practical situations one usually has to fold the answers with the appropriate neutrino flux. In Sect.~\ref{general_tensors} we introduce the general electroweak leptonic and hadronic tensors. We use the notation already employed in studies of electron scattering (see, for instance, \cite{DR86,RD89}), while in Sects.~\ref{leptonic_tensor} and \ref{hadronic_tensor} the details of these two tensors are further developed. In Sect.~\ref{contraction} the tensor contractions and semi-inclusive cross section are presented and finally, in Sect.~\ref{conclusions} we summarize the results of this study and indicate where we are presently applying the formalism using specific models.

% Kinematics
\section{Kinematics \label{kinematics}}

To describe the kinematics of the particles involved in the process we indicate four-vectors with capital letters such as $A^{\mu}=(A^{0},A^{1},A^{2},A^{3})$ and three-vectors with boldface lower-case letters such as $\mathbf{a}$, with their magnitudes in normal-faced font $a=|\mathbf{a}|$; the metric used, as in \cite{BjD64},
yields $A\cdot B=A_{\mu }B^{\mu }=A^{0}B^{0}-\mathbf{a}\cdot
\mathbf{b}$ (repeated indices summed).

\begin{figure}
\begin{center}
\includegraphics[width=0.8\textwidth]{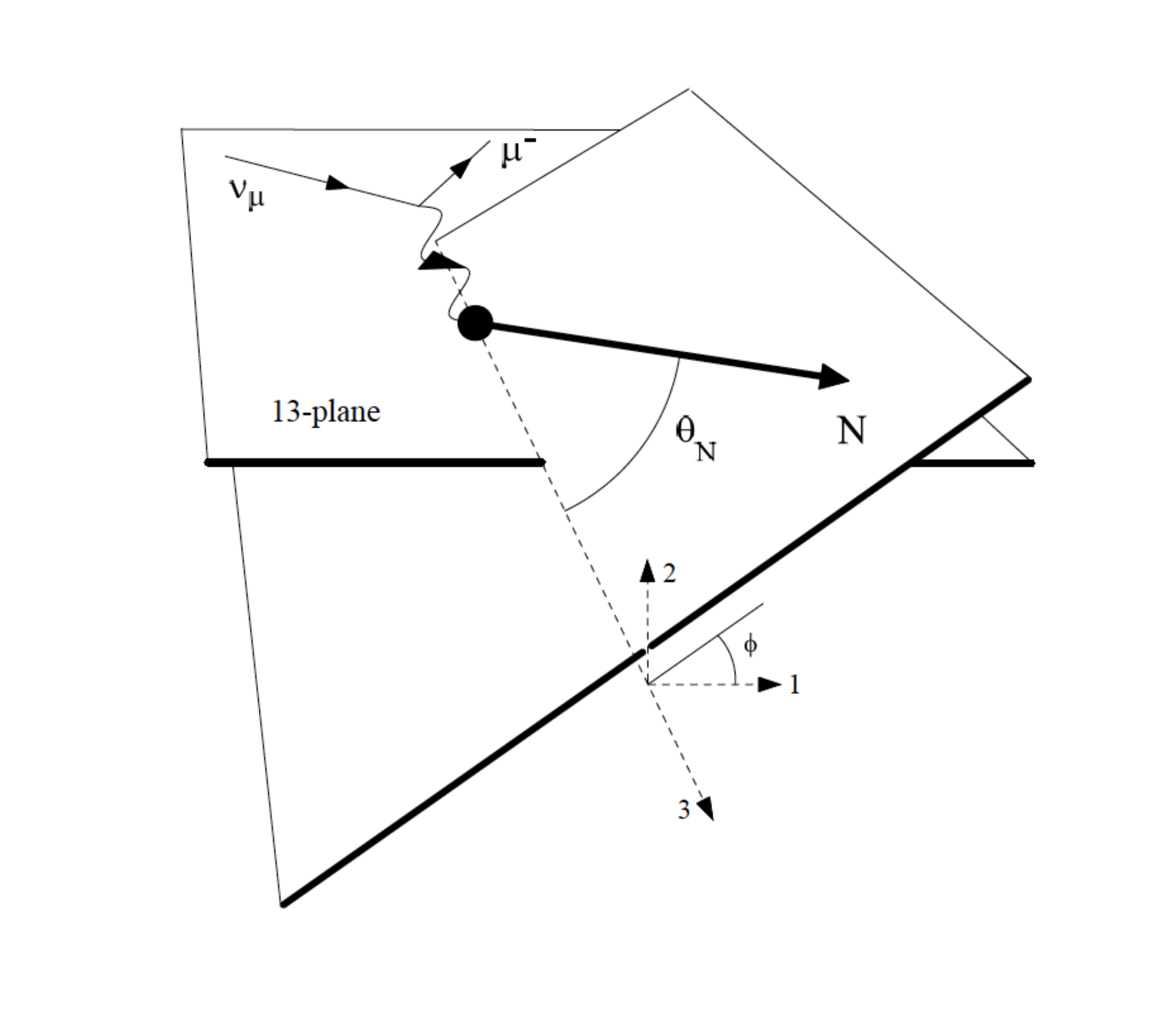}
\caption{Kinematics for semi-inclusive neutrino-nucleus reactions. \label{figure_kinematics}}
\end{center}
\end{figure}

The incident neutrino (or antineutrino) carries four-momentum $K^{\mu }=(\varepsilon ,\mathbf{k})$, where $\varepsilon =\sqrt{k^{2}+m^{2}}$ is the total energy, $\mathbf{k}$ is the three-momentum and $m$ is the mass. The outgoing charged lepton has four-momentum $K^{\prime \mu }=(\varepsilon^{\prime },\mathbf{k}^{\prime })$ and mass $m'$. The space-like four-momentum of the boson exchanged with the nuclear target is $Q^{\mu }=(\omega ,\mathbf{q})$, with $-Q^2=|Q^2|=q^2-\omega^2\ge 0$. We assume the three-momentum $\mathbf{q}$ to be along the $3$-axis so that the incoming and outgoing leptons define the $13$-plane (see Fig. \ref{figure_kinematics}). By defining the lepton scattering angle $\theta$ (\textit{i.e.}, the angle between $\mathbf{k}$ and $\mathbf{k}^\prime $), the components of the incident and outgoing leptons and exchanged boson four-momenta can be written as:
\begin{equation}
\begin{array}{lll}
K^{0} = \varepsilon & K^{\prime 0} = \varepsilon^{\prime}  & Q^{0} =\varepsilon -\varepsilon^{\prime} = \omega \\
K^{1} = \frac{1}{q}kk^{\prime }\sin \theta & K^{\prime 1} = \frac{1}{q}kk^{\prime }\sin \theta & Q^{1} = 0 \\
K^{2} = 0 & K^{\prime 2} = 0 & Q^{2} = 0 \\
K^{3} = \frac{1}{q}k\left( k-k^{\prime }\cos \theta \right) \quad\quad & K^{\prime 3} = -\frac{1}{q}k^{\prime} \left(k^{\prime}-k\cos \theta \right) \quad\quad & Q^{3} =\sqrt{k^{2}+k^{\prime 2}-2kk^{\prime }\cos \theta} = q
\end{array}
\end{equation}

In the laboratory system the incoming nuclear target with mass $M^{0}_{A}$ carries four-momentum $P_{A}^{\mu}=(M^{0}_{A},0,0,0)$. We assume that the final hadronic state consists of a stripped nucleon and the remaining daughter nucleus with four-momenta $P_{N}^{\mu }=(E_{N},\mathbf{p}_{N})$ and $P_{A-1}^{\mu }=(E_{A-1},\mathbf{p}_{A-1})$ respectively. This $A-1$ daughter system may be in its ground state, in some discrete excited state, may be an $A-2$ granddaughter nucleus plus a nucleon, {\it etc.}, and has invariant mass $W_{A-1}$. The only assumption so far is that one nucleon is presumed to be detected and so only final states with one or more nucleons, at least one being of the appropriate flavor, are being considered (see also below). Using the coordinate system introduced above, where $\mathbf{q}$ lies along the $z$-axis and the leptons lie in the $13$-plane, the total four-momentum in the hadronic vertex is
\begin{eqnarray}
P_{tot}^{\mu } &\equiv &Q^{\mu }+P_{A}^{\mu }=P_{N}^{\mu }+P_{A-1}^{\mu } = (P_{tot}^{0},0,0,P_{tot}^{3})  \label{eqh2a} \\
P_{tot}^{0} &=&M^{0}_{A}+\omega =E_{N}+E_{A-1}\equiv E  \label{eqh2b} \\
\mathbf{p}_{tot} &=&P_{tot}^{3}\mathbf{u}_{3}=q\mathbf{\mathbf{u}_{3}=q}=
\mathbf{p}_{N}+\mathbf{p}_{A-1}.  \label{eqh3}
\end{eqnarray}
Writing out the components of the products' four-momenta one has
\begin{equation}
\begin{array}{ll}
P_{N}^{0}=E_{N} & P_{A-1}^{0}=E_{A-1} \\
P_{N}^{1}=p_{N}\sin \theta _{N}\cos \phi \qquad & P_{A-1}^{1}=-p_{A-1}\sin \theta
_{A-1}\cos \phi \\
P_{N}^{2}=p_{N}\sin \theta _{N}\sin \phi \qquad & P_{A-1}^{2}=-p_{A-1}\sin \theta
_{A-1}\sin \phi \\
P_{N}^{3}=p_{N}\cos \theta _{N} & P_{A-1}^{3}=p_{A-1}\cos \theta _{A-1}
\end{array}
\label{eqh4}
\end{equation}
where $\theta_N$ and $\theta_{A-1}$ are the angles of the hadronic products with respect to the $3$-axis (direction of ${\bf q}$), and $\phi$ is the angle between the plane defined by the nucleon momentum $\boldmath{p}_N$ and the momentum transfer $\boldmath{q}$ and the leptonic ($13$) plane. From Eqs.~(\ref{eqh3}) and (\ref{eqh4}) we have that
\begin{eqnarray}\label{eqh5}
\sin \theta _{A-1}=\frac{1}{p_{A-1}}p_{N}\sin \theta _{N} \qquad \qquad \cos \theta _{A-1}=\frac{1}{p_{A-1}}\left( q-p_{N}\cos \theta _{N}\right)
\end{eqnarray}
and from conservation of energy, Eq.~(\ref{eqh2b}), one has $E_{A-1}=E-E_{N}$, where both products are on-shell, {\it i.e.}, $E_{N}=\sqrt{p_{N}^2+m_{N}^2}$ and $E_{A-1}=\sqrt{p_{A-1}^2+W_{A-1}^2}$, where $W_{A-1}$, as said above, is the invariant mass of the $A-1$ daughter system.

Having set up the basic form for the semi-inclusive cross section, let us next consider the problem in more detail by discussing the general kinematical variables to be used when studying $X(\nu_{\ell} ,\ell ^{-}N)$ and $X(\bar{\nu}_{\ell} ,\ell ^{+}N)$ reactions in context with previous studies of $X(e,e^{\prime }N)$ reactions. We have seen above that the cross section depends on a limited set of kinematic variables. The leptonic variables are those discussed above. The hadronic variables, in contrast, are best transformed into other variables when treating semi-inclusive scattering from nuclei. We shall see in the following section that the dependences on the azimuthal angle $\phi$ can be made explicit using the general Lorentz structure of the hadronic tensor and so we can leave that variable aside. We have the momentum transfer $q$ and energy transfer $\omega$ from the leptonic side via the exchange of a single $W^{\pm}$, and so we can use this pair or equivalently $Q^2$ and $\nu$ in other notation, or $Q^2$ and $x\equiv |Q^2|/2m_N \nu$ in still other notation. That leaves us with $p_N$ and $\theta_N$ which are more conveniently transformed in two new variables. While these sets of dynamical variables are, of course, completely usable and indeed natural from an experimental point of view, we shall see in the following that alternative sets are more convenient when studying the specifics of the cross section in the regime of quasifree scattering.

From three-momentum conservation one has
\begin{equation}
\mathbf{p}_{A-1}=\mathbf{q}-\mathbf{p}_{N}\equiv -\mathbf{p,}  \label{eqsc1}
\end{equation}
where $\mathbf{p}$ is minus the missing momentum $\mathbf{p}_{m}$, so that
the daughter energy becomes $E_{A-1}=\sqrt{W_{A-1}^{2}+p^{2}}$. This is
completely general and, in particular is not dictated by any specific model for the reaction. Clearly this momentum merely characterizes the split in momentum flow between the
detected nucleon and the unobserved daughter nucleus. From energy conservation and using the
three-momentum conservation relation, one has
\begin{eqnarray}
M_{A}^{0}+\omega = E_{N}+E_{A-1} = \sqrt{q^{2}+p^{2}+2qp\cos \theta _{pq}+m_{N}^{2}}+\sqrt{p^{2}+W_{A-1}^{2}} \label{ener_cons}
\end{eqnarray}
with $\theta _{pq}$ being the angle between $\mathbf{p}$ and $\mathbf{q}$.
Next we need some energy variable to characterize the degree of excitation
of the daughter nucleus. A natural choice is the excitation energy in the
rest frame of the recoiling daughter nucleus, $E^{\ast }\equiv W_{A-1}-W_{A-1}^{0}\geq 0$,
where $W_{A-1}$ includes the internal excitation energy of the $A-1$ system
while $W_{A-1}^{0}$ is the smallest possible invariant mass of the $A-1$ and will be the ground state rest mass of this system $M^0_{A-1}$ in most cases. By construction $E^{\ast }$ is greater than or equal to zero --- and equal to zero when the daughter nucleus is left in its ground state. Using this one
can obtain the so-called missing energy
\begin{equation}
E_{m}=m_{N}+W_{A-1}-M_{A}^{0}=E_{s}+E^{\ast }  \label{eqsc5a}
\end{equation}
where $E_s = m_N + W^0_{A-1}-M^0_A$ is the separation energy (or \textquotedblleft $Q$--value\textquotedblright ), another commonly used energy in the problem is defined as the minimum energy needed to separate the nucleus $A$ into a nucleon and the residual nucleus $A-1$ in its ground state.
As we shall see below, we could now use $(E^{\ast },p)$ or $(E_{m},p_{m})$
in place of $(E_{N},\theta _{N})$, although it may be shown that still
another choice for the energy is preferable for certain purposes than $E^{\ast}$, namely
\begin{eqnarray}
\mathcal{E} &\equiv &E_{A-1}-E_{A-1}^{0}\geq 0  \label{eqsc6}
\end{eqnarray}
where as before $E_{A-1}=\sqrt{W_{A-1}^{2}+p^{2}}$ and now also $E_{A-1}^{0}=\sqrt{{W_{A-1}^{0}}^{2}+p^{2}}$. This quantity does not differ much from the excitation energy $E^{\ast}$ for $p << W^0_{A-1}$, which is typically the case; let us call it \textquotedblleft daughter energy difference\textquotedblright, in contrast to the \textquotedblleft daughter excitation energy\textquotedblright\ $E^{\ast }$.

Overall energy conservation yields an equation for $\mathcal{E}$ in terms of $q$, $\omega $, $p$ and the angle $\theta _{pq}$:
\begin{equation}
\mathcal{E} = \omega - E_s +m_{N} - \sqrt{m_{N}^{2}+p^{2}+q^{2}+2pq\cos\theta _{pq}}-\sqrt{{W^0_{A-1}}^2+p^2}+ W^0_{A-1},  \label{eqsc14}
\end{equation}
Thus there are clear relationships between the sets $(E_{N},\theta _{N})$ and $(p,\theta )$ and hence $(\mathcal{E},p)$. Instead of the first set, we shall now use the last set as a pair of dynamical variables.

With these preliminaries in hand let us discuss the characteristic landscape of the coincidence semi-inclusive cross section as a function of $\mathcal{E}$ and $p$ for fixed $q$ and $\omega $ (and of course fixed $\theta $ and $\phi$). We have not yet required that the kinematic relationships discussed above should be satisfied, and when we do so, we find that only specific regions are accessible. Noting that Eq.~(\ref{eqsc14}) yields a curve of $\mathcal{E}$ versus $p$ in the $(\mathcal{E},p)$--plane for each choice of $\theta _{pq}$, let us see what constraint the requirement that $-1\leq \cos \theta _{pq}\leq +1$ imposes on the kinematics.

First, consider \textquotedblleft $\omega $ small\textquotedblright\ (to be specified completely below) and plot the trajectory when $\cos \theta_{pq}=-1$. A curve rising from negative $\mathcal{E}$ to intersect $\mathcal{E}=0$ at $p=p_{min}>0$ which peaks at some value of $p$ and then falls to intersect $\mathcal{E}=0$ again, this time at $p=p_{max}>p_{min}$, is generally obtained. All physically allowable values of $\mathcal{E}$ and $p$ must lie below this curve and, of course, above $\mathcal{E}=0$. To obtain the other extreme, $\cos \theta _{pq}=+1$, one can simply replace $p$ by $-p$ in Eq.~(\ref{eqsc14}); the physically allowable values of $\mathcal{E}$ and $p$ must lie above this curve. For \textquotedblleft $\omega$ small\textquotedblright , no physically allowable values at all occur near the latter curve and the physical region is completely defined by the $\cos\theta _{pq}=-1$ curve and $\mathcal{E}=0$. Following past work \cite{Day90} we shall call the minimum value of momentum $p_{min}\equiv -y$ and the maximum value $p_{max}\equiv +Y$. The formal definition of\textquotedblleft $\omega $ small\textquotedblright\ then becomes \textquotedblleft $y<0$\textquotedblright . We can set $\mathcal{E}=0$ in Eq.~(\ref{eqsc14}) and solve for $y$ and $Y$, yielding
\begin{eqnarray}
y(q,\omega ) &=&\frac{1}{W_A^{2}}\left[ (M_{A}^{0}+\omega )\sqrt{\Lambda^{2}-{W_{A-1}^{0}}^{2}W_A^{2}}-q\Lambda \right]   \label{eqsc15} \\
Y(q,\omega ) &=&\frac{1}{W_A^{2}}\left[ (M_{A}^{0}+\omega )\sqrt{\Lambda^{2}-{W_{A-1}^{0}}^{2}W_A^{2}}+q\Lambda \right]   \label{eqsc16}
\end{eqnarray}
with
\begin{eqnarray}
W_A &=&\sqrt{(M_{A}^{0}+\omega )^{2}-q^{2}}  \label{eqsc17} \\
\Lambda  &=&\frac{1}{2}\left( W_A^{2}+{W_{A-1}^{0}}^{2}-m_{N}^{2}\right) .
\label{eqsc18}
\end{eqnarray}
A useful relationship is the following:
\begin{equation}
M_{A}^{0}+\omega =\sqrt{(q+y)^{2}+m_{N}^{2}}+\sqrt{y^{2}+{W_{A-1}^{0}}^{2}}.
\label{eqsc19}
\end{equation}
Noting that --- approximately --- the quasielastic peak occurs at the
kinematical point where $y=0$, it is useful to use Eq.~(\ref{eqsc19}) to
define
\begin{equation}
\omega _{QE}\equiv \omega (y=0)\equiv \left\{ \sqrt{q^{2}+m_{N}^{2}}
-m_{N}\right\} +E_{s}=|Q_{QE}^{2}|/2m_{N}+E_{s}.  \label{eqsc23}
\end{equation}
Accordingly,  \textquotedblleft $\omega $ small\textquotedblright\
corresponds to $y<0,$ namely to $\omega <\omega _{QE}$. Finally, the
equation for the upper boundary of the allowed region (\textit{i.e.,\/}
corresponding to $\cos \theta _{pq}=-1$) is given by
\begin{equation}
\mathcal{E}_{-}=\sqrt{m_{N}^{2}+(q+y)^{2}}-\sqrt{m_{N}^{2}+(q-p)^{2}}+\sqrt{{W_{A-1}^{0}}^{2}+y^{2}}-\sqrt{{W_{A-1}^{0}}^{2}+p^{2}}.  \label{eqsc25}
\end{equation}
When the momentum transfer becomes very large one can show that this goes to
the \emph{finite} asymptotic limit
\begin{equation}
\mathcal{E}_{-}\underset{q\rightarrow \infty }{\longrightarrow }\mathcal{E}_{-}^{\infty }=y+p-\left[ \sqrt{{W_{A-1}^{0}}^{2}+p^{2}}-\sqrt{{W_{A-1}^{0}}^{2}+y^{2}}\right] .  \label{eqsc27}
\end{equation}
Henceforth, instead of the sets $\{q,\omega ,E_{N},\theta _{N}\}$ or $\{Q^{2},Q\cdot P_{A},P_{N}\cdot P_{A},Q\cdot P_{N}\}$ we shall use the set $\{q,y,\mathcal{E},p\}$ to characterize the general two-arm coincidence
cross section. In particular, the response functions to be introduced later on are all
functions of these four variables together with $\phi$.

\begin{figure}
\begin{center}
\includegraphics[width=1\textwidth]{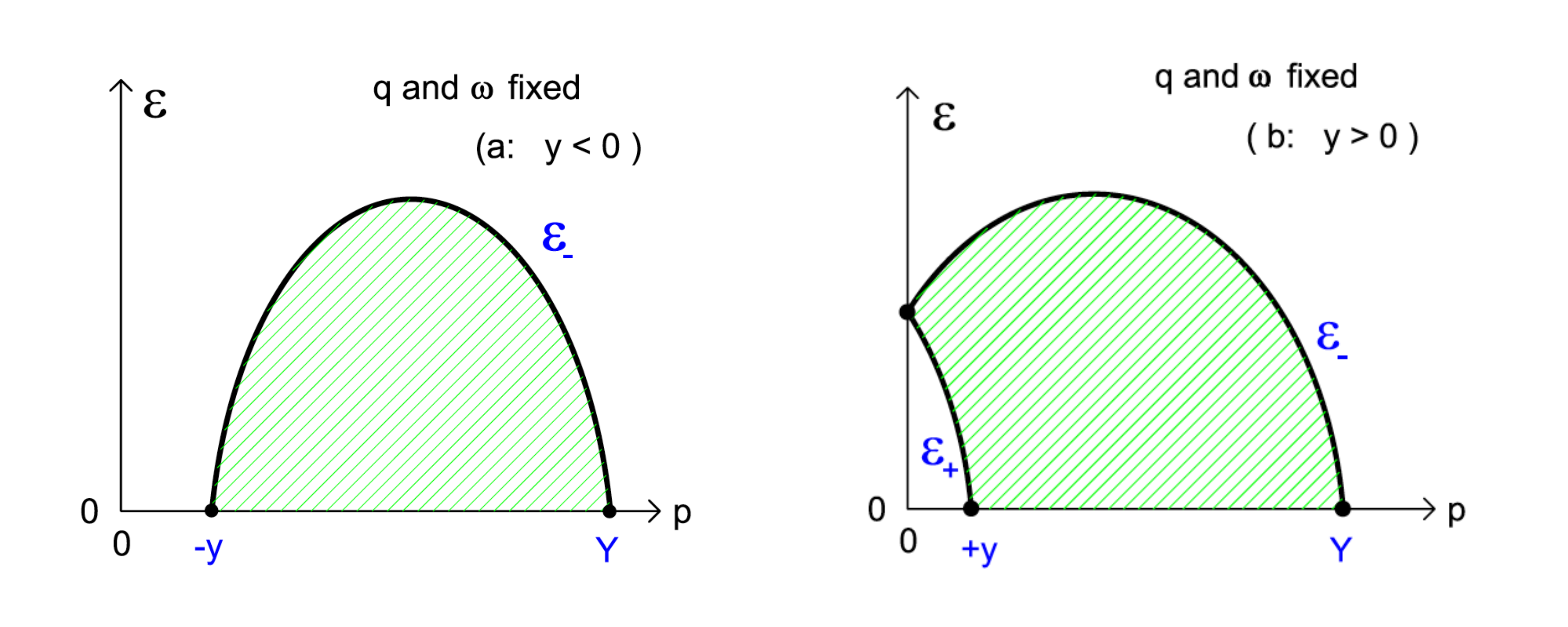}
\caption{(Color online) Planes defined by the daughter energy difference $\mathcal{E}$ and the missing momentum $p$, showing the allowed region for semi-inclusive neutrino-nucleus scattering processes. Left: (a)  For $y<0$, {\it i.e.}, $\omega$ below the quasielastic peak. Right (b): For $y>0$, {\it i.e.}, $\omega$ above the quasielastic peak. \label{fig_kin_y}}
\end{center}
\end{figure}

In Fig. \ref{fig_kin_y} (a) are shown families of curves of $\mathcal{E}_{-}$ versus $p$ for specific values of $q$ and $y<0$. The physical regions lie below these curves and above $\mathcal{E}=0$ for the chosen kinematics. Clearly, by imposing these kinematic constraints on the semi-inclusive cross section it is possible to see what features of the dynamics are or are not accessible in the $y<0$ region. Note that even when $q\rightarrow \infty $ only a limited part of the dynamical landscape is accessible. Also note that inclusive scattering corresponds to integrating over the entire accessible region for $q$ and $y$ (or equivalently $\omega $) fixed, and summing over the allowed particle species ($N=p$ and $n$), and correcting for double-counting by subtracting the cross section where both a proton and a neutron are detected in coincidence with the charged lepton.

These developments can be extended rather easily to the \textquotedblleft $\omega $ large\textquotedblright\ region, which becomes equivalent to $y>0$ and hence to $\omega >\omega _{QE}$.  Again the curves of $\mathcal{E}$ versus $p$ when $\cos \theta _{pq}=\pm 1$ define boundaries. The $\cos\theta _{pq}=-1$ curve (namely, $\mathcal{E}=\mathcal{E}_{-}$ above) is much as before, except that now $p_{min}$ is negative and so $y\equiv -p_{min}$ is positive. Reflecting $p\rightarrow -p$ to obtain the $\cos \theta _{pq}=+1 $ curve from the $\cos \theta _{pq}=-1$ curve as before now yields a nontrivial result: the physically allowable region must lie below the $\cos\theta _{pq}=-1$ curve and above the $\cos \theta _{pq}=+1$ curve, and since the latter lies in the quadrant where $\mathcal{E}\geq 0$ and $p\geq 0$, this provides a new boundary, namely $\mathcal{E}_{-}$ obtained from Eq.~(\ref{eqsc25}) by changing $p$ to $-p$. In Fig. \ref{fig_kin_y} (b) results similar to those
in Fig. \ref{fig_kin_y} (a) are shown, except now for $y\geq 0$. The physically accessible region in each case lies \emph{above }the lines extending from $p=y$ to the $\mathcal{E}$--axis and \emph{below} the curves extending from the $\mathcal{E}$--axis to peak at some value of $p$ and fall again, eventually intersecting the $\mathcal{E}=0$ line at $p_{max}=Y$. Again we see that only specific parts of the semi-inclusive cross section are accessible for these kinematics.

The merit of transforming to the $({\cal E},p)$ variables is that these are best suited to characterizing the nuclear dynamics. The semi-inclusive cross section, as studied to some extent via reactions, has its most important contributions lying at relatively small values of
${\cal E}$, where one typically finds distributions as functions of $p$ that reflect the shell structure of the specific nucleus being studied. For instance, in a simple shell model of the nucleus one sees features that reflect the knockout of nucleons from the valence shell, the next-to-valence shell, {\it etc.} These fall relatively rapidly with increasing $p$. Unfortunately, however, such simple models are not adequate and one also requires overall suppression of these ``momentum distributions'' by factors of typically 30\% via the so-called spectroscopic factors. Also from studies one knows that some of this ``missing strength'' is moved to higher values of ${\cal E}$, partially through standard nuclear interactions which make both initial and final nuclear states complicated. Said another way, the states involved are undoubtedly not simple single Slater determinants. Also, the NN interaction has both long- and short-range contributions, and especially the latter can promote strength to higher ${\cal E}$ and $p$. Something like 20-30\% of the strength is known to reside in this part of the landscape, although the actual amounts are not very well determined. In between the two regions one has other likely issues to deal with, namely the fact that there are several open channels to be considered and these can conspire via channel-coupling to produce the true final many-body state. An example is when a nucleon is ejected from a deep-lying shell model state: for typical kinematics it is also possible to have two or more nucleons ejected and these channels can couple, yielding a very complex situation. Such issues are very hard to treat, especially in a relativistic context as is required for typical studies of neutrino reactions.

% Electroweak Tensors
\section{General Electroweak Tensors \label{general_tensors}}

The cross section takes on its characteristic form involving the contraction of two second-rank Lorentz tensors, $d\sigma \sim \eta _{\mu \nu }W^{\mu \nu }$, corresponding to the leptonic and the hadronic contributions which are thus factorized and dealt with independently. The leptonic tensor is defined as
\begin{equation}
\eta _{\mu \nu }\equiv 2mm^{\prime} \:\overline{\sum_{if}}j_{\mu }^{\ast }j_{\nu },
\label{def_lep_tensor_general}
\end{equation}
where a factor $4mm^{\prime}$ (merged here with an additional factor 1/2) has been included to compensate spinor norms later on, the lepton masses being kept finite until the end of our developments. Its hadronic counterpart is
\begin{equation}
W^{\mu \nu } \equiv \overline{\sum_{if}} J_{fi}^{\mu \ast }(\mathbf{q})J_{fi}^{\nu }(\mathbf{q}),  \label{eq4-2}
\end{equation}
where the operations $\overline{\sum }_{if}$ in the two cases correspond to sums and averages over the appropriate sets of leptonic quantum numbers (the helicities, in fact) or hadron quantum numbers
(helicities or spins, \textit{etc.}) and integration over all unobserved particles in the final state of the $A-1$ system for hadrons. It proves useful to decompose both leptonic and hadronic tensors into pieces which are symmetric ($s$) or antisymmetric ($a$) under index interchange $\mu \leftrightarrow \nu$, since in contracting them no symmetric-antisymmetric cross-terms are allowed. Both tensors can thus be decomposed as   $\eta _{\mu \nu } = \eta _{\mu \nu }^{s}+\eta _{\mu \nu }^{a}$ and $W^{\mu \nu } = W_{s}^{\mu \nu }+W_{a}^{\mu \nu }$, where the terms are defined as
\begin{equation}
\begin{array}{ll}
 \eta _{\mu \nu }^{s} = \frac{1}{2}(\eta_{\mu \nu } + \eta_{\nu \mu }) \:; \qquad & \qquad \eta _{\mu \nu }^{a} = \frac{1}{2}(\eta_{\mu \nu } - \eta_{\nu \mu }) \:; \\
 W^{\mu \nu }_{s} = \frac{1}{2}(W^{\mu \nu } + W^{\nu \mu }) \:; \qquad & \qquad W^{\mu \nu }_{a} = \frac{1}{2}(W^{\mu \nu } - W^{\nu \mu }) \:.
\end{array}
 \end{equation}
Clearly one has that $\eta _{\mu \mu }^{s} = \eta _{\mu \mu }$ and $W_{s}^{\mu \mu } = W^{\mu \mu }$, whereas $\eta _{\mu \mu }^{a} = W_{a}^{\mu \mu } = 0$ (no summation over $\mu $ implied in these expressions). In addition, since each tensor is proportional to the bilinear combinations of the electroweak currents in the forms $\eta _{\mu \nu }\sim j_{\mu }^{*}j_{\nu }$ and $W^{\mu \nu }\sim J^{\mu *}J^{\nu }$, one has that $\eta _{\mu \nu }^{*} = \eta _{\nu \mu }$ and $W^{\mu \nu *} = W^{\nu \mu }$, and thus that
\begin{equation}
\begin{array}{ll}
\eta _{\mu \nu }^{s} = \text{Re}\:\eta _{\mu \nu } \:;  \qquad & \qquad \eta _{\mu \nu }^{a} = i\:\text{Im}\:\:\eta _{\mu \nu }  \:; \\
W_{s}^{\mu \nu } = \text{Re}\:W^{\mu \nu }  \:; \qquad & \qquad W_{a}^{\mu \nu } = i\:\text{Im}\:W^{\mu \nu }  \:.\label{eqcon9}
\end{array}
\end{equation}

Let us begin by defining the following (real) symmetric (no prime) and antisymmetric (prime) hadronic response functions:
\begin{eqnarray}
W^{CC} &\equiv &\:\text{Re}\:W^{00}=W^{00}  \label{wcc} \\
W^{CL} &\equiv &2\:\text{Re}\:W^{03}=2\:W_{s}^{03}  \label{wcl} \\
W^{LL} &\equiv &\:\text{Re}\:W^{33}=W^{33}  \label{wll} \\
W^{T} &\equiv &\:\text{Re}\:W^{22}+\:\text{Re}\:W^{11}=W^{22}+W^{11}
\label{wt} \\
W^{TT} &\equiv &\:\text{Re}\:W^{22}-\:\text{Re}\:W^{11}=W^{22}-W^{11} \label{wtt} \\
W^{TC} &\equiv &2\sqrt{2}\:\text{Re}\:W^{01}=2\sqrt{2}\:W_{s}^{01} \label{wtc} \\
W^{TL} &\equiv &2\sqrt{2}\:\text{Re}\:W^{31}=2\sqrt{2}\:W_{s}^{31} \label{wtl} \\
W^{\underline{TT}} &\equiv &2\:\text{Re}\:W^{12}=2\:W_{s}^{12} \label{wttu} \\
W^{\underline{TC}} &\equiv &2\sqrt{2}\:\text{Re}\:W^{02}=2\sqrt{2}\:W_{s}^{02} \label{wtcu} \\
W^{\underline{TL}} &\equiv &2\sqrt{2}\:\text{Re}\:W^{32}=2\sqrt{2}\:W_{s}^{32} \label{wtlu} \\
W^{T^{\prime }} &\equiv & -2\:\text{Im}\:W^{12} = 2\:iW_{a}^{12}  \label{wtp}\\
W^{TC^{\prime }} &\equiv &-2\sqrt{2}\:\text{Im}\:W^{02}=2\sqrt{2}\:iW_{a}^{02} \label{wtcp} \\
W^{TL^{\prime }} &\equiv &-2\sqrt{2}\:\text{Im}\:W^{32}=2\sqrt{2}\:iW_{a}^{32}  \label{wtlp} \\
W^{\underline{CL}^{\prime }} &\equiv &  \:\text{Im}\:W^{03} = iW_{a}^{03}\label{wclup}\\
W^{\underline{TC}^{\prime }} &\equiv &2\sqrt{2}\:\text{Im}\:W^{01}=-2\sqrt{2}\:iW_{a}^{01}  \label{wtcup} \\
W^{\underline{TL}^{\prime }} &\equiv &2\sqrt{2}\:\text{Im}\:W^{31}=-2\sqrt{2}\:iW_{a}^{31} \label{wtlup}
\end{eqnarray}

Here $C$ refers to charge (the $\mu =0$) projection, $L$ refers to longitudinal (momentum transfer direction, $\mu =3$) projection and $T$ refers to transverse ($\mu =1,2$) projections. Concerning the latter, the meaning of the combinations used above can be elucidated by introducing the spherical components of the transverse projections of the hadronic current, defined as
\begin{eqnarray}
J^{(+1)} = -\frac{1}{\sqrt{2}}\left( J^{1} + iJ^{2}\right) \:;\quad\quad J^{(-1)} = \frac{1}{\sqrt{2}}\left( J^{1} - iJ^{2}\right) \:;\quad\quad J^{(0)} = J^3
\end{eqnarray}
or inversely:
\begin{eqnarray}
J^{1} = -\frac{1}{\sqrt{2}}\left( J^{(+1)} - J^{(-1)}\right) \:;\quad\quad J^{2} = \frac{i}{\sqrt{2}}\left( J^{(+1)} + J^{(-1)}\right) \:;\quad\quad J^3 = J^{(0)}
\end{eqnarray}
With these definitions, and using the notation $W^{(mm')}$ for the spherical vector components ($m,m'=\{+1,-1,0\}$) of the hadronic tensor, one can rewrite the responses that contain transverse projections as:
\begin{eqnarray}
W^{T} &\equiv& W^{(+1+1)}+W^{(-1-1)}  \label{eqcon21-1} \\
W^{TT} &\equiv& 2\:\text{Re} \:W^{(+1-1)}  \label{eqcon21-2} \\
W^{TL} &\equiv& -2\:\text{Re} \left( W^{(0+1)}-W^{(0-1)}\right)  \label{eqcon28-2} \\
W^{\underline{TT}} &\equiv& 2\:\text{Im}\:W^{(+1-1)}  \label{eqcon23-2}\\
W^{\underline{TL}} &\equiv& -2\:\text{Im}\left( W^{(0+1)}+W^{(0-1)}\right)  \label{eqcon28-8}\\
W^{T^{\prime }} &\equiv& W^{(+1+1)}-W^{(-1-1)}  \label{eqcon23-1} \\
W^{TL^{\prime }} &\equiv& -2\:\text{Re}\left( W^{(0+1)}+W^{(0-1)}\right)  \label{eqcon28-4} \\
W^{\underline{TL}^{\prime }} &\equiv& -2\:\text{Im}\left( W^{(0+1)}-W^{(0-1)}\right)  \label{eqcon28-6}\\
\end{eqnarray}
It is thus clear that the $T$ response, being an incoherent sum of circularly (or linearly) polarized responses, is the unpolarized transverse response, whereas the $TT$ response contains the information needed to specify the linear polarization information (more clearly seen in Eq.~(\ref{wtt})). The $T^{\prime }$ response, on the other hand, gives the additional information needed together with the $T$ response to specify the circular polarization.

Equivalently to the hadronic case, the corresponding symmetric (no prime) and antisymmetric (prime) leptonic quantities may be defined:
\begin{eqnarray}
v_{0}\widehat{V}_{CC} &\equiv &\:\text{Re}\:\eta_{00}=\eta_{00} \label{vcc} \\
v_{0}\widehat{V}_{CL} &\equiv &\:\text{Re}\:\eta_{03}=\eta_{03}^{s} \label{vcl} \\
v_{0}\widehat{V}_{LL} &\equiv &\:\text{Re}\:\eta_{33}=\eta_{33} \label{vll} \\
v_{0}\widehat{V}_{T} &\equiv &\frac{1}{2}(\text{Re}\:\eta_{22}+\:\text{Re}\:\eta_{11})=\frac{1}{2}(\eta_{22}+\eta_{11})  \label{vt} \\
v_{0}\widehat{V}_{TT} &\equiv &\frac{1}{2}(\text{Re}\:\eta_{22}-\:\text{Re}\:\eta_{11})=\frac{1}{2}(\eta_{22}-\eta _{11}) \label{vtt} \\
v_{0}\widehat{V}_{TC} &\equiv &\frac{1}{\sqrt{2}}\:\text{Re}\:\eta_{01} = \frac{1}{\sqrt{2}}\:\eta^{s}_{01}\label{vtc} \\
v_{0}\widehat{V}_{TL} &\equiv &\frac{1}{\sqrt{2}}\:\text{Re}\:\eta_{31} = \frac{1}{\sqrt{2}}\:\eta^{s}_{31} \label{vtl} \\
v_{0}\widehat{V}_{\underline{TT}} &\equiv &\text{Re}\:\eta_{12} =  \eta^{s}_{12} \label{vttu} \\
v_{0}\widehat{V}_{\underline{TC}} &\equiv &\frac{1}{\sqrt{2}}\:\text{Re}\:\eta_{02} = \frac{1}{\sqrt{2}}\:\eta^{s}_{02} \label{vtcu} \\
v_{0}\widehat{V}_{\underline{TL}} &\equiv &\frac{1}{\sqrt{2}}\:\text{Re}\:\eta_{32} = \frac{1}{\sqrt{2}}\:\eta^{s}_{32} \label{vtlu} \\
v_{0}\widehat{V}_{T^{\prime }} &\equiv &\:\text{Im}\:\eta_{12} = -i\eta^{a}_{12} \label{vtp} \\
v_{0}\widehat{V}_{TC^{\prime }} &\equiv &\frac{1}{\sqrt{2}}\:\text{Im}\:\eta_{02} = -\frac{1}{\sqrt{2}}\:i\eta^{a}_{02} \label{vtcp} \\
v_{0}\widehat{V}_{TL^{\prime }} &\equiv &\frac{1}{\sqrt{2}}\:\text{Im}\:\eta_{32} = -\frac{1}{\sqrt{2}}\:i\eta^{a}_{32} \label{vtlp} \\
v_{0}\widehat{V}_{\underline{CL}^{\prime }} &\equiv &-\:\text{Im}\:\eta_{03} = i\eta^{a}_{03}  \label{vclup}\\
v_{0}\widehat{V}_{\underline{TC}^{\prime }} &\equiv &-\frac{1}{\sqrt{2}}\:\text{Im}\:\eta_{01} = \frac{1}{\sqrt{2}}\:i\eta^{a}_{01} \label{vtcup} \\
v_{0}\widehat{V}_{\underline{TL}^{\prime }} &\equiv &-\frac{1}{\sqrt{2}}\:\text{Im}\:\eta_{31} = \frac{1}{\sqrt{2}}\:i\eta^{a}_{31} \label{vtlup}
\end{eqnarray}
where the overall factor $v_{0}$ is defined as
\begin{eqnarray}\label{v0}
v_{0} \equiv (\varepsilon+\varepsilon^{\prime})^2 - q^2 .
\end{eqnarray}

The results found here are completely general; they are simply a convenient rewriting of the original components of the leptonic and hadronic tensors where the projections along the momentum transfer direction ($L$) and transverse to it provide the organizing principle.

% Leptonic Tensor
\subsection{Leptonic Tensor \label{leptonic_tensor}}

From definition in Eq.~(\ref{def_lep_tensor_general}) and employing the conventions of \cite{BjD64} we form the general leptonic tensor involving neutrinos and negatively charged leptons --- later it is straightforward to extend the results to include antineutrinos and positively charged leptons:
\begin{eqnarray}
\eta _{\mu \nu }(K^{\prime },K) = mm^{\prime}\sum_{s,s^{\prime}} \bar{u}(K,s)\:(a_V\gamma_{\mu}+a_A\gamma_{\mu}\gamma_{5})\:u(K^{\prime},s^{\prime})\:\bar{u}(K^{\prime},s^{\prime})\:(a_V\gamma_{\nu}+a_A\gamma_{\nu}\gamma_{5})\:u(K,s) \nonumber \\
\label{def_lep_tensor}
\end{eqnarray}
which includes sum over final spin states and average over initial spin states, the latter implying a factor 1/2. In the standard model the charged-current vector and axial coupling constants take the values $a_V = 1$ and  $a_A = -1$, which yields the usual form of the vertex $\gamma_{\mu}\:(1-\gamma_{5})$. Upon eliminating the spinors using traces one finds:
\begin{eqnarray}
\eta _{\mu \nu }(K^{\prime },K) &\equiv& \frac{1}{4}\left\{ \mathrm{Tr}[a_{V}\gamma _{\mu }+a_{A}\gamma _{\mu }\gamma _{5}]\:(K'\!\!\!\!\!\! /\ +m^{\prime }) \:[a_{V}\gamma _{\nu }+a_{A}\gamma _{\nu }\gamma _{5}]\:(K\!\!\!\!\!\! /\ +m)\right\}  \label{eq10} \\\nonumber
&= & \frac{1}{4}\left\{ a_{V}^{2}\mathrm{Tr}\left[ \gamma _{\mu }(K'\!\!\!\!\!\! /\ +m^{\prime })\gamma _{\nu }(K\!\!\!\!\!\thinspace /+m)\right] _{(1)} +a_{A}^{2}\mathrm{Tr}\left[ \gamma _{\mu }\gamma _{5}(K'\!\!\!\!\!\! /\ +m^{\prime })\gamma _{\nu }\gamma _{5}(K\!\!\!\!\!\thinspace / +m)\right] _{(2)}  \right. \\\nonumber
&& \left. +a_{V}a_{A}\left( \mathrm{Tr}\left[ \gamma _{\mu }(K'\!\!\!\!\!\! /\ +m^{\prime })\gamma _{\nu }\gamma _{5}(K\!\!\!\!\!\thinspace / +m)\right] _{(3)} +\mathrm{Tr}\left[ \gamma _{\mu }\gamma _{5}(K'\!\!\!\!\!\! /\ +m^{\prime })\gamma _{\nu }(K\!\!\!\!\!\thinspace / +m)\right] _{(4)}\right) \right\}.  \label{eq11}
\end{eqnarray}
The traces can then be expressed as:
\begin{eqnarray}
\frac{1}{4}\mathrm{Tr}\left[ {}\right] _{(1)} &=&K_{\mu }K_{\nu }^{\prime}+K_{\mu }^{\prime }K_{\nu }-g_{\mu \nu }\left( K\cdot K^{\prime}-mm^{\prime }\right)  \label{eq12} \\
\frac{1}{4}\mathrm{Tr}\left[ {}\right] _{(2)} &=&K_{\mu }K_{\nu }^{\prime}+K_{\mu }^{\prime }K_{\nu }-g_{\mu \nu }\left( K\cdot K^{\prime}+mm^{\prime }\right)  \label{eq13} \\
\frac{1}{4}\mathrm{Tr}\left[ {}\right] _{(3)} &=&\frac{1}{4}\mathrm{Tr}\left[{}\right] _{(4)}=-i\varepsilon _{\mu \nu \alpha \beta }K^{\alpha }K^{\prime \beta }  \label{eq14}
\end{eqnarray}

Cases (1) and (2) are symmetric under interchange of $\mu $ with $\nu $, while cases (3) and (4) (the VA-interference terms) are antisymmetric. Note that if studying reactions with an incident or outgoing massless leptons ($m=0$ or $m^{\prime }=0$) then cases (1) and (2) yield the same answer.

We introduce the following definitions:
\begin{eqnarray}
&& \nu  \equiv \frac{\omega }{q}  \label{nu} \\
&& \rho  \equiv \frac{|Q^{2}|}{q^{2}}=1-\nu ^{2}  \:;\qquad\qquad \rho^{\prime } \equiv \frac{q}{\varepsilon +\varepsilon^{\prime}} \label{rho}\\
&& \delta  \equiv \frac{m}{\sqrt{|Q^{2}|}} \:;\qquad\qquad \delta ^{\prime }\equiv \frac{m^{\prime }}{\sqrt{|Q^{2}|}}  \label{delta}\\
&& \tan^{2}\widetilde{\theta }/2 = \frac{|Q^2|}{v_0} = \frac{\rho\rho^{\prime 2}}{1-\rho^{\prime 2}} \label{tan}.
\end{eqnarray}
In terms of the angle $\widetilde{\theta}$ the quantities $Q^2$ and $v_0$ (the latter defined in Eq.~(\ref{v0})) can be written as
\begin{eqnarray}
&& Q^2 = -4\:\varepsilon \:\varepsilon^{\prime} \:\sin^2\widetilde{\theta}/2 \\
&& v_0 = 4\:\varepsilon \:\varepsilon^{\prime} \:\cos^2\widetilde{\theta}/2.
\end{eqnarray}
Using the previous definitions the components of the leptonic tensor as defined in Eqs.~(\ref{vcc}--\ref{vtlup}) give rise to the following expressions
\begin{eqnarray}
\widehat{V}_{CC} & = & \frac{1}{2} \:\left\{\left(a_{V}^{2}+a_{A}^{2}\right) -\left[ a_{V}^{2}\left( \delta -\delta ^{\prime}\right) ^{2}+a_{A}^{2}\left( \delta +\delta ^{\prime }\right) ^{2}\right] \tan ^{2}\widetilde{\theta }/2 \right\} \label{vcc2} \\
\widehat{V}_{CL} & = & -\frac{1}{2} \:\left(a_{V}^{2}+a_{A}^{2}\right) \left[ \nu -\frac{1}{\rho ^{\prime }}\left(\delta ^{2}-\delta ^{\prime 2}\right) \tan ^{2}\widetilde{\theta }/2\right] \label{vcl2} \\\nonumber
\widehat{V}_{LL} & = & \frac{1}{2} \:\left\{\left(a_{V}^{2}+a_{A}^{2}\right) \left[ \nu ^{2}-\frac{1}{\rho ^{\prime }}\left(2\nu -\rho \rho ^{\prime }\left( \delta ^{2}-\delta ^{\prime 2}\right)\right) \left( \delta ^{2}-\delta ^{\prime 2}\right) \tan ^{2}\widetilde{\theta }/2\right]  \right. \\
&& \left. +\left[ a_{V}^{2}\left( \delta -\delta ^{\prime }\right)^{2}+a_{A}^{2}\left( \delta +\delta ^{\prime }\right) ^{2}\right] \tan ^{2}\widetilde{\theta }/2  \right\} \label{vll2} \\\nonumber
\widehat{V}_{T} & = & \frac{1}{2} \:\left(a_{V}^{2}+a_{A}^{2}\right) \left\{ \left[ \frac{1}{2}\rho +\tan ^{2}\widetilde{\theta }/2\right] \right.   \\\nonumber
&&\left. +\left( \frac{\nu }{\rho ^{\prime }}\left( \delta^{2}-\delta^{\prime 2}\right) -\frac{1}{2}\rho \left( \delta^{2}-\delta ^{\prime 2}\right) ^{2}\right) \tan ^{2}\widetilde{\theta }/2\right\}   \label{vt}\\
&&-\left( a_{V}^{2}-a_{A}^{2}\right) \delta \delta^{\prime }\tan^{2}\widetilde{\theta }/2   \\\nonumber
\widehat{V}_{TT} & = & \frac{1}{2} \:\left(a_{V}^{2}+a_{A}^{2}\right) \left\{ -\frac{1}{2}\rho \right.    \\
&&\left. +\left[ \left( \delta ^{2}+\delta ^{\prime 2}\right) -\frac{\nu }{\rho ^{\prime }}\left( \delta ^{2}-\delta ^{\prime 2}\right) +\frac{1}{2}\rho \left( \delta ^{2}-\delta ^{\prime 2}\right) ^{2}\right] \tan ^{2}\widetilde{\theta }/2\right\}   \label{vtt2} \\\nonumber
\widehat{V}_{TC} & = & -\frac{1}{2} \:\left( a_{V}^{2}+a_{A}^{2}\right) \frac{1}{\rho ^{\prime }}\tan \widetilde{\theta }/2  \\
&&\times \left( \frac{1}{2} -\frac{1}{\rho }\left[ \left( \delta ^{2}+\delta ^{\prime 2}\right) -\frac{\nu }{\rho ^{\prime }}\left( \delta ^{2}-\delta ^{\prime 2}\right) +\frac{1}{2}\rho \left( \delta ^{2}-\delta ^{\prime 2}\right) ^{2}\right] \tan ^{2}\widetilde{\theta }/2\right) ^{1/2}  \label{vtc2} \\
\widehat{V}_{TL} & = & -\left( \nu -\rho \rho ^{\prime }\left( \delta ^{2}-\delta^{\prime 2}\right) \right) \widehat{V}_{TC}  \label{vtl2} \\
\widehat{V}_{\underline{TT}} & = & 0  \label{vttu2} \\
\widehat{V}_{\underline{TC}} & = & 0  \label{vtcu2} \\
\widehat{V}_{\underline{TL}} & = & 0  \label{vtlu2} \\
\widehat{V}_{T^{\prime }} & = & a_{V}a_{A}\frac{1}{\rho ^{\prime }}\left( 1+\nu \rho ^{\prime }\left(\delta ^{2}-\delta ^{\prime 2}\right) \right) \tan ^{2}\widetilde{\theta }/2 \label{vtp2} \\\nonumber
\widehat{V}_{TC^{\prime }} & = & -a_{V}a_{A}\tan \widetilde{\theta }/2  \\
&&\times \left[ \frac{1}{2}-\frac{1}{\rho }\left[ \left( \delta ^{2}+\delta^{\prime 2}\right) -\frac{\nu }{\rho ^{\prime }}\left( \delta ^{2}-\delta^{\prime 2}\right) +\frac{1}{2}\rho \left( \delta ^{2}-\delta ^{\prime2}\right) ^{2}\right] \tan ^{2}\widetilde{\theta }/2\right] ^{1/2}   \label{vtcp2} \\
\widehat{V}_{TL^{\prime }} & = & -\nu \widehat{V}_{TC^{\prime }}  \label{vtlp2} \\
\widehat{V}_{\underline{CL}^{\prime }} & = & 0  \label{vclup2} \\
\widehat{V}_{\underline{TC}^{\prime }} & = & 0  \label{vtcup2} \\
\widehat{V}_{\underline{TL}^{\prime }} & = & 0  \label{vtlup2}
\end{eqnarray}

Within these 16 factors, 10 of them are symmetric and 6 are antisymmetric. Under the conditions in this work 6 of them vanish, namely the ones with underlined subscript (see \cite{RD89} for processes where they do not); the rest reduce to the following expressions in the extreme relativistic limit (ERL), defined as $\widehat{V}_{K}\xrightarrow{ERL} \frac{1}{2}\left(a_{V}^{2}+a_{A}^{2}\right) v_{K}$ for the symmetric ones (no prime) and as $\widehat{V}_{K^{\prime}}\xrightarrow{ERL} a_{V}a_{A}\: v_{K^{\prime}}$ for the antisymmetric ones (prime):
\begin{eqnarray}
v_{CC} &=&1  \label{eq46} \\
v_{CL} &=&-\nu  \label{eq47} \\
v_{LL} &=&\nu ^{2}  \label{eq48} \\
v_{T\:} &=&\frac{1}{2}\rho +\tan ^{2}\theta /2  \label{eq49} \\
v_{TT} &=&-\frac{1}{2}\rho  \label{eq50} \\
v_{TC} &=&-\frac{1}{\sqrt{2}\rho ^{\prime }}\tan \theta /2  \label{eq51} \\
v_{TL} &=&-\nu v_{TC}  \label{eq52} \\
v_{T^{\prime }} &=&\tan \theta /2\sqrt{\rho +\tan ^{2}\theta /2} \\
v_{TC^{\prime }} &=&-\frac{1}{\sqrt{2}}\tan \theta /2 \\
v_{TL^{\prime }} &=&-\nu v_{TC^{\prime }}
\label{eq55}
\end{eqnarray}

It is worth noticing that the following combination is useful when discussing conserved vector
current (CVC) terms:
\begin{eqnarray}\nonumber
\widehat{V}_{L} &\equiv &\widehat{V}_{CC}+2\nu \widehat{V}_{CL}+\nu ^{2}
\widehat{V}_{LL}  \label{eq45a} \\\nonumber
&=& \frac{1}{2}\:\left( a_{V}^{2}+a_{A}^{2}\right) \left\{ \rho ^{2}+\nu \rho \left[ \frac{2}{\rho ^{\prime }}+\nu \left( \delta ^{2}-\delta ^{\prime 2}\right) \right]
\left( \delta ^{2}-\delta ^{\prime 2}\right) \tan ^{2}\widetilde{\theta }/2\right\}  \\
&&-\frac{1}{2}\:\left[ a_{V}^{2}\left( \delta -\delta ^{\prime }\right)
^{2}+a_{A}^{2}\left( \delta +\delta ^{\prime }\right) ^{2}\right] \rho \tan
^{2}\widetilde{\theta }/2  \label{eq45b},
\end{eqnarray}
whose corresponding ERL factor is $v_{L} = \rho^{2}$. Also, the $T$ and $TT$ terms are simply related:
\begin{equation}
\widehat{V}_{T}+\widehat{V}_{TT}= \frac{1}{2}\:\left\{ \left( a_{V}^{2}+a_{A}^{2}\right) +
\left[ a_{V}^{2}\left( \delta -\delta ^{\prime }\right) ^{2}+a_{A}^{2}\left(
\delta +\delta ^{\prime }\right) ^{2}\right] \right\} \tan ^{2}\widetilde{\theta }/2 . \label{eq45}
\end{equation}

Finally, one can easily complete the leptonic developments by going to the start and replacing the $u$-spinor by $v$-spinors so that the leptonic tensor for anti-particles can be obtained. The final result is that upon contracting the leptonic and hadronic tensors (see Sect.~\ref{contraction}) the VV and AA terms are as above, while the VA interference changes sign.

% Hadronic tensor
\subsection{Hadronic Tensor \label{hadronic_tensor}}

Among the various components of the hadronic tensor defined above only some of them occur, which can be deduced from the general developments of the hadronic tensor as it is constructed from the available
four-momenta. The reaction of interest here is semi-inclusive scattering where, as we have seen in Sect.~\ref{kinematics}, at the hadronic vertex one has incoming momentum transfer $Q^{\mu }$ and the nuclear target momentum $P_{A}^{\mu }$. In the final state one has the momentum of the detected nucleon $P_{N}^{\mu }$ together with the residual nucleus' momentum which can be eliminated using four-momentum conservation: $P_{A-1}^{\mu }=Q^{\mu }+P_{A}^{\mu }-P_{N}^{\mu }$. Six invariants can be constructed:
\begin{equation}
\begin{array}{lll}
I_{1} \equiv Q^{2} \:;\qquad & I_{2} \equiv Q\cdot P_{A} \:;\qquad & I_{3} \equiv Q\cdot P_{N} \:; \\
I_{4} \equiv P_{A}\cdot P_{N} \:;\qquad & I_{5} \equiv P_{A}^{2}={M^{0}_{A}}^{2} \:;\qquad & I_{6} \equiv P_{N}^{2}=m_{N}^{2} \:,
\end{array}
\end{equation}
of which the first four are dynamical variables, whereas the last two are fixed by the target nucleus and nucleon masses. Accordingly all invariant structure functions depend on the four dynamical invariants $I_{i},$ $i=1,\ldots,4$. They can be expressed as:
\begin{eqnarray}
I_{1} &=&\omega ^{2}-q^{2}<0 \\
I_{2} &=&M^{0}_{A}\omega \\
I_{3} &=&\omega E_{N}-qp_{N}\cos \theta _{N} \\
I_{4} &=&M^{0}_{A}E_{N}
\end{eqnarray}

Next one can write symmetric and antisymmetric hadronic tensors as functions
of the three independent four-momenta $Q^{\mu }$, $P_{A}^{\mu }$ and $P_{N}^{\mu }$. In fact, it proves to be more convenient to introduce projected four-momenta to replace the last two, namely,
\begin{eqnarray}
U^{\mu } &\equiv &\frac{1}{M^{0}_{A}}\left[ P_{A}^{\mu }-\left( \frac{Q\cdot
P_{A}}{Q^{2}}\right) Q^{\mu }\right]   \label{eqr14} \\
V^{\mu } &\equiv &\frac{1}{M_{N}}\left[ P_{N}^{\mu }-\left( \frac{Q\cdot
P_{N}}{Q^{2}}\right) Q^{\mu }\right] ,  \label{eqr15}
\end{eqnarray}
where then $Q\cdot U=Q\cdot V=0$. Also, to keep the dimensions consistent in
the developments below let us introduce a dimensionless four-momentum transfer
\begin{equation}
\widetilde{Q}^{\mu }\equiv \frac{Q^{\mu }}{\sqrt{|Q^{2}|}}.  \label{eqr15a}
\end{equation}

The symmetric hadronic tensor may then be written
\begin{eqnarray}\nonumber
W_{s}^{\mu \nu } &=&X_{1}g^{\mu \nu }+X_{2}\widetilde{Q}^{\mu }\widetilde{Q}^{\nu }+X_{3}U^{\mu }U^{\nu }+X_{4}(\widetilde{Q}^{\mu }U^{\nu }+U^{\mu }\widetilde{Q}^{\nu })  \\
&&+X_{5}V^{\mu }V^{\nu }+X_{6}(\widetilde{Q}^{\mu }V^{\nu }+V^{\mu }\widetilde{Q}^{\nu })+X_{7}\left( U^{\mu }V^{\nu }+V^{\mu }U^{\nu }\right) ,
\label{eqr16}
\end{eqnarray}
where $X_{i}$, $i=1\ldots 7$ are invariant functions of the invariants discussed above. These seven types of terms arise from VV and AA contributions. Likewise the antisymmetric tensor can be constructed from the
basic four-momenta
\begin{eqnarray}\nonumber
W_{a}^{\mu \nu } &=&i\left\{ Y_{1}(\widetilde{Q}^{\mu }U^{\nu }-U^{\mu }\widetilde{Q}^{\nu }) +Y_{2}(\widetilde{Q}^{\mu }V^{\nu }-V^{\mu }\widetilde{Q}^{\nu})+Y_{3}(U^{\mu }V^{\nu }-V^{\mu }U^{\nu })  \right.\label{eqr17} \\
&&\left. +Z_{1}\varepsilon ^{\mu \nu \alpha \beta }\widetilde{Q}_{\alpha}U_{\beta }+Z_{2}\varepsilon ^{\mu \nu \alpha \beta }\widetilde{Q}_{\alpha}V_{\beta }+Z_{3}\varepsilon ^{\mu \nu \alpha \beta }U_{\alpha }V_{\beta}\right\}  ,
\end{eqnarray}
where again $Y_{i}$ and $Z_{i}$, $i=1\ldots 3$ are invariant functions of the invariants above. The terms having no $\varepsilon^{\mu \nu \alpha \beta }$, namely the $Y_i$ terms (as well as the $X_i$ terms, as said above), arise from VV and AA contributions, whereas those with $\varepsilon^{\mu \nu \alpha \beta }$, namely the $Z_i$ terms, come from VA interferences. Note that for inclusive scattering where one does not have $V^{\mu }$ as a building block only terms of the $X_{1}$, $X_{2}$, $X_{3}$, $X_{4}$, $Y_{1}$ and $Z_{1}$ type can occur.

For a conserved vector current (CVC) situation such as here for the VV terms or for purely polar-vector
electron scattering the continuity equation in momentum space requires that
\begin{equation}
Q_{\mu }\left( W_{s}^{\mu \nu }\right) _{VV}=Q_{\mu }\left( W_{a}^{\mu \nu
}\right) _{VV}=0.  \label{eqr18}
\end{equation}
For the symmetric tensor this contraction removes the terms with $X_{3}$, $X_{5},Y_{3}$, $Z_{1}$, leaving the conditions
\begin{eqnarray}
&& \left( -X_{1}^{VV}+X_{2}^{VV}\right) \widetilde{Q}^{\nu }+X_{4}^{VV}U^{\nu}+X_{6}^{VV}V^{\nu } = 0  \label{eqr19} \\
&& Y_{1}^{VV}U^{\nu }+Y_{2}^{VV}V^{\nu } = 0,  \label{eqr20}
\end{eqnarray}
where no terms with $\varepsilon ^{\mu \nu \alpha \beta }$ can occur in a VV
situation, \textit{i.e.,} $Z_{1}^{VV}=Z_{2}^{VV}=Z_{3}^{VV}=0$, as noted
above. Since the basic four-momenta are linearly independent of each other the coefficients above must all be independently zero, namely $X_{1}^{VV}-X_{2}^{VV}=X_{4}^{VV}=X_{6}^{VV}=Y_{1}^{VV}=Y_{2}^{VV}=0$. Accordingly, one has
\begin{eqnarray}\nonumber
\left( W_{s}^{\mu \nu }\right) _{VV} &=&X_{1}^{VV}\left[ g^{\mu \nu }-\frac{Q^{\mu }Q^{\nu }}{Q^{2}}\right] +X_{3}^{VV}U^{\mu }U^{\nu }  \\
&&+X_{5}^{VV}V^{\mu }V^{\nu }+X_{7}^{VV}\left( U^{\mu }V^{\nu }+V^{\mu}U^{\nu }\right)  \label{eqr21} \\
\left( W_{a}^{\mu \nu }\right) _{VV} &=&Y_{3}^{VV}(U^{\mu }V^{\nu }-V^{\mu}U^{\nu }).  \label{eqr22}
\end{eqnarray}
For instance, in semi-inclusive electron scattering the symmetric terms lead to the standard $L$, $T$, $TL$ and $TT$ responses, while the antisymmetric term which becomes accessible with polarized electron scattering yields the $TL^{\prime}$ response, the so-called 5th response \cite{DR86,RD89}. For the other cases, the AA and VA responses, there is no further simplification in general. The resulting number of contributions of each type is summarized in Table \ref{responses_numbers} for semi-inclusive and for inclusive scattering, the latter arising from integrating the semi-inclusive contributions. For the semi-inclusive case of interest here, they form the functions $X_i, Y_i, Z_i$ as follows:
\begin{equation}
\begin{array}{ll}\label{had_resp_terms}
X_{1} = X_{1}^{VV}+X_{1}^{AA}  \qquad & \qquad Y_{1} = Y_{1}^{AA} \\
X_{2} = X_{1}^{VV}+X_{2}^{AA}  \qquad & \qquad Y_{2} = Y_{2}^{AA} \\
X_{3} = X_{3}^{VV}+X_{3}^{AA}  \qquad & \qquad Y_{3} = Y_{3}^{VV}+Y_{3}^{AA} \\
X_{4} = X_{4}^{AA}  \qquad & \qquad Z_{1} = Z_{1}^{VA} \\
X_{5} = X_{5}^{VV}+X_{5}^{AA}  \qquad & \qquad  Z_{2} = Z_{2}^{VA} \\
X_{6} = X_{6}^{AA}  \qquad & \qquad Z_{3} = Z_{3}^{VA} \\
X_{7} = X_{7}^{VV}+X_{7}^{AA} \qquad & \qquad
\end{array}
\end{equation}

\hspace{2in}
\begin{table}
\begin{center}
\begin{tabular}{|c|c|c|c|c|}
\hline$\quad$
 & \multicolumn{2}{|c|}{Semi-inclusive} & \multicolumn{2}{|c|}{Inclusive} \\ \hline
$\mathrm{Type}$ & $\quad$ Sym $\quad$ & $\quad$ A-sym $\quad$ & $\quad$ Sym $\quad$ & $\quad$ A-sym $\quad$ \\ \hline
$\quad$ VV $\quad$ & $\quad$ 4 $\quad$ & $\quad$ 1 $\quad$ & $\quad$ 2 $\quad$ & $\quad$ 0 $\quad$ \\ \hline
$\quad$ AA $\quad$ & $\quad$ 7 $\quad$ & $\quad$ 3 $\quad$ & $\quad$ 4 $\quad$ & $\quad$ 1 $\quad$ \\ \hline
$\quad$ VA $\quad$ & $\quad$ 0 $\quad$ & $\quad$ 3 $\quad$ & $\quad$ 0 $\quad$ & $\quad$ 1 $\quad$ \\ \hline
\end{tabular}
\end{center}
\caption{Number of electroweak responses in semi-inclusive and inclusive processes, classified according to their properties under spatial inversion (VV, AA, and VA) and index interchange (symmetric and antisymmetric). \label{responses_numbers}}
\end{table}

Upon using the kinematic variables in the laboratory system discussed in Sect.~\ref{kinematics}, in particular Eqs.~(\ref{nu}, \ref{rho}), together with the following definitions:
\begin{eqnarray}
\eta _{T} &\equiv &\frac{p_{N}}{m_{N}}\sin \theta _{N}  \label{eqr15c} \\
H &\equiv &\frac{1}{m_{N}}\left[ E_{N}-\nu p_{N}\cos \theta _{N}\right] ,
\label{eqr15d}
\end{eqnarray}
the hadronic response functions defined in Sect.~\ref{general_tensors} can be written as
\begin{eqnarray}\nonumber
W_{s}^{CC} &=&\frac{1}{\rho ^{2}}\left\{ \rho ^{2}X_{1}+\rho \nu
^{2}X_{2}+X_{3}+2\sqrt{\rho}\nu X_{4}\right.  \\
&&\left. +H^{2}X_{5}+2\sqrt{\rho}\nu HX_{6}+2HX_{7}\right\}  \label{eqr23}\\\nonumber
W_{s}^{CL} &=&\frac{2\nu}{\rho^{2}}\left\{ \rho X_{2}+X_{3}+\sqrt{\rho }(\frac{1}{\nu }+\nu )X_{4}\right.  \\
&&\left. +H^{2}X_{5}+\sqrt{\rho}(\frac{1}{\nu }+\nu )HX_{6}+2HX_{7}\right\}\label{eqr24} \\\nonumber
W_{s}^{LL} &=&\frac{1}{\rho^{2}}\left\{ -\rho ^{2}X_{1}+\rho X_{2}+\nu^{2}X_{3}+2\sqrt{\rho }\nu X_{4}\right.  \\
&&\left. +\nu ^{2}H^{2}X_{5}+2\sqrt{\rho }\nu HX_{6}+2\nu ^{2}HX_{7}\right\} \label{eqr25} \\
W_{s}^{T} &=&-2X_{1}+X_{5}\eta _{T}^{2}  \label{eqr26} \\
W_{s}^{TT} &=&-X_{5}\eta _{T}^{2}\cos 2\phi  \label{eqr27} \\
W_{s}^{TC} &=&\frac{2\sqrt{2}}{\rho }\eta _{T}\left\{ HX_{5}+\sqrt{\rho }\nu X_{6}+X_{7}\right\} \cos \phi \label{eqr29} \\
W_{s}^{TL} &=&\frac{2\sqrt{2}}{\rho }\eta _{T}\left\{ \nu HX_{5}+\sqrt{\rho }
X_{6}+\nu X_{7}\right\} \cos \phi  \label{eqr31} \\
W_{s}^{\underline{TT}} &=&X_{5}\eta _{T}^{2}\sin 2\phi  \label{eqr28} \\
W_{s}^{\underline{TC}} &=&\frac{2\sqrt{2}}{\rho}\eta _{T}\left\{ HX_{5}+\sqrt{\rho }\nu X_{6}+X_{7}\right\}\sin \phi  \label{eqr30} \\
W_{s}^{\underline{TL}} &=&\frac{2\sqrt{2}}{\rho}\eta _{T}\left\{ \nu HX_{5}+\sqrt{\rho }X_{6}+\nu X_{7}\right\} \sin \phi \label{eqr32} \\
W_{a}^{T^{\prime }} &=&\frac{1}{\sqrt{\rho }}\left\{ Z_{1}+HZ_{2}\right\} \label{eqr34} \\
W_{a}^{TC^{\prime }} &=&\frac{2\sqrt{2}}{\rho }\eta _{T}\left\{ -\left( \sqrt{\rho }\nu Y_{2}+Y_{3}\right) \sin \phi +\left( \sqrt{\rho }Z_{2}+\nu Z_{3}\right) \cos \phi \right\}  \label{eqr35} \\
W_{a}^{TL^{\prime }} &=&\frac{2\sqrt{2}}{\rho }\eta _{T}\left\{ -\left(\sqrt{\rho }Y_{2}+\nu Y_{3}\right) \sin \phi +\left( \sqrt{\rho }\nu Z_{2}+Z_{3}\right) \cos \phi \right\}  \label{eqr37} \\
W_{a}^{\underline{CL}^{\prime }} &=&-\frac{1}{\sqrt{\rho }}\left\{Y_{1}+HY_{2}\right\}  \label{eqr33} \\
W_{a}^{\underline{TC}^{\prime }} &=&\frac{2\sqrt{2}}{\rho }\eta _{T}\left\{\left( \sqrt{\rho }\nu Y_{2}+Y_{3}\right) \cos \phi +\left( \sqrt{\rho } Z_{2}+\nu Z_{3}\right) \sin \phi \right\}  \label{eqr36} \\
W_{a}^{\underline{TL}^{\prime }} &=&\frac{2\sqrt{2}}{\rho }\eta _{T}\left\{\left( \sqrt{\rho }Y_{2}+\nu Y_{3}\right) \cos \phi +\left( \sqrt{\rho }\nu Z_{2}+Z_{3}\right) \sin \phi \right\}   \label{eqr38}
\end{eqnarray}

Note how the explicit dependence on the azimuthal angle $\phi $ emerges: one has pairs of symmetric contributions, namely $TT\leftrightarrow \underline{TT}$, $TC\leftrightarrow \underline{TC}$, and $TL\leftrightarrow \underline{TL}$, where a cosine is replaced by a sine, as well as pairs of antisymmetric contributions, namely, $TC^{\prime}\leftrightarrow \underline{TC^{\prime}}$ and $TL^{\prime}\leftrightarrow \underline{TL^{\prime}}$, where a rotation is involved. Also note that, while these constitute the complete set of semi-inclusive responses, in fact none of the underlined cases enter when combined with the leptonic factors obtained above, since the latter are all zero (see Eqs.~(\ref{vcc2}--\ref{vtlup2})).

% Tensors contraction and cross section
\section{Contraction of tensors and cross section}\label{contraction}

The contraction of the leptonic and the hadronic tensors arises from the application of standard Feynman rules to the evaluation of the cross section of the process under study here; it is an invariant, taking the same form in the laboratory, in the center-of-momentum, or in any other system of reference. As mentioned in Sect.~\ref{general_tensors}, the symmetric and the antisymmetric components of the leptonic and the hadronic tensors can be contracted separately since no cross-terms are allowed:
\begin{eqnarray}
v_0 \:\mathcal{F}_{\chi}^2 \equiv \eta _{\mu \nu }W^{\mu \nu } = \eta _{\mu \nu }^{s}W_{s}^{\mu \nu } + \chi \:\eta _{\mu \nu }^{a}W_{a}^{\mu \nu } \:,
\end{eqnarray}
where $\chi = 1$ for incident neutrinos, as obtained in Section \ref{leptonic_tensor}, and $\chi = -1$ for antineutrinos, as can be easily shown with the same formalism but using antiparticle spinors $v$ in Eq.~(\ref{def_lep_tensor}). In Cartesian components the symmetric and the antisymmetric contractions above yield
\begin{eqnarray}\nonumber
\eta _{\mu \nu }^{s}W_{s}^{\mu \nu } &=& \eta _{00}^{s}W_{s}^{00}+2\eta_{03}^{s}W_{s}^{03}+\eta _{33}^{s}W_{s}^{33} +\eta _{11}^{s}W_{s}^{11}+\eta _{22}^{s}W_{s}^{22}  \\
&& +2\eta _{01}^{s}W_{s}^{01}+2\eta _{31}^{s}W_{s}^{31}+2\eta _{02}^{s}W_{s}^{02}+2\eta _{32}^{s}W_{s}^{32}  +2\eta _{12}^{s}W_{s}^{12} \label{eqcon7-1} \\ \nonumber
\eta _{\mu \nu }^{a}W_{a}^{\mu \nu } &=& 2\eta _{03}^{a}W_{a}^{03} +2\eta _{01}^{a}W_{a}^{01}+2\eta_{31}^{a}W_{a}^{31}  \\
&& +2\eta _{02}^{a}W_{a}^{02}+2\eta _{32}^{a}W_{a}^{32} +2\eta _{12}^{a}W_{a}^{12},  \label{eqcon7-2}
\end{eqnarray}
which, according to the developments of Sect.~\ref{general_tensors}, can be expressed as
\begin{eqnarray}\nonumber
\eta _{\mu \nu }^{s}W_{s}^{\mu \nu } &=& \text{Re}\:\eta _{00}\:\text{Re}\:W^{00}+2
\:\text{Re}\:\eta _{03}\:\text{Re}\:W^{03}+\:\text{Re}\:\eta _{33}\:\text{Re}\:W^{33} \\\nonumber
&&+\:\text{Re}\:\eta _{11}\:\text{Re}\:W^{11}+\:\text{Re}\:\eta _{22}\:\text{Re}\:W^{22}
+2\:\text{Re}\:\eta _{01}\:\text{Re}\:W^{01} \\\nonumber
&&+2\:\text{Re}\:\eta _{31}\:\text{Re}\:W^{31} +2\:\text{Re}\:\eta _{02}\:\text{Re}\:W^{02}+2\:\text{Re}\:\eta _{32}\:\text{Re}\:W^{32} \\
&&+2\:\text{Re}\:\eta _{12}\:\text{Re}\:W^{12}  \label{eqcon10-1}
\end{eqnarray}
\begin{eqnarray}\nonumber
-\eta _{\mu \nu }^{a}W_{a}^{\mu \nu } &=& 2\:\text{Im}\:\eta _{03}\:\text{Im}\:W^{03} +2\:\text{Im}\:\eta _{01}\:\text{Im}\:W^{01}+2\:\text{Im}\:\eta _{31}\:\text{Im}\:W^{31} \\
&&+2\:\text{Im}\:\eta _{02}\:\text{Im}\:W^{02}+2\:\text{Im}\:\eta _{32}\:\text{Im}\:W^{32} +2\:\text{Im}\:\eta _{12}\:\text{Im}\:W^{12}.  \label{eqcon10-2}
\end{eqnarray}

Finally, in terms of projections with respect to the momentum transfer direction the contractions read
\begin{eqnarray}\nonumber
\eta _{\mu \nu }^{s}W_{s}^{\mu \nu } &=& v_{0}\left\{ \left[ \widehat{V}_{CC}W^{CC}+\widehat{V}_{CL}W^{CL}+\widehat{V}_{LL}W^{LL}\right. \right.  \\\nonumber
&& \left. +\widehat{V}_{T}W^{T}+\widehat{V}_{TT}W^{TT}  +\widehat{V}_{TC}W^{TC}+\widehat{V}_{TL}W^{TL}\right]  \\
&& \left. +\left[ \widehat{V}_{\underline{TT}}W^{\underline{TT}}+\widehat{V}_{\underline{TC}}W^{\underline{TC}}+\widehat{V}_{\underline{TL}}W^{\underline{TL}}\right] \right\} ,  \label{eqcon30}
\end{eqnarray}
\begin{eqnarray}\nonumber
\eta _{\mu \nu }^{a}W_{a}^{\mu \nu } &=& v_{0}\left\{ \left[ \widehat{V}_{T^{\prime }}W^{T^{\prime }}+\widehat{V}_{TC^{\prime }}W^{TC^{\prime }}+\widehat{V}_{TL^{\prime }}W^{TL^{\prime }}\right] \right.  \\
&& \left. +\left[ \widehat{V}_{\underline{CL}^{\prime }}W^{\underline{CL}^{\prime }}+\widehat{V}_{\underline{TC}^{\prime }}W^{\underline{TC}^{\prime}}+\widehat{V}_{\underline{TL}^{\prime }}W^{\underline{TL}^{\prime }}\right]
\right\} ,  \label{eqcon31}
\end{eqnarray}
where the hadronic responses contain all the VV, AA, and VA terms applicable to each of them, as shown in Eqs.~(\ref{had_resp_terms}).

In any of the above representations the symmetric contraction involves 10 terms and the antisymmetric one involves 6 terms, for an expected total of 16 terms. From the tensor contractions above the matrix element of the process is (see definition of the leptonic tensor in Eq.~(\ref{def_lep_tensor})):
\begin{equation}
|\mathcal{M}_{\chi}|^2 = \frac{G^2\:\cos^2 \theta_c \:v_0}{2mm'} \:\mathcal{F}_{\chi}^{2} \:,
\end{equation}
where $G = 1.166 \times 10^{-5}$ GeV$^{-2}$ is the coupling constant of the weak interaction, $\cos \theta_c=0.974$ with $\theta_c$ the Cabibbo angle accounting for the misalignment between the strong and the weak hadronic eigenstates, $v_0$ was defined in Eq.~(\ref{v0}), and, as said above, $\chi=+1$ for neutrino and $\chi=-1$ for antineutrino scattering.

We then evaluate the coincidence cross section of the processes $^{A}X(\nu_{\ell} ,\ell^- N)^{A-1}Y$ or $^{A}X(\bar{\nu}_{\ell} ,\ell^+ N)^{A-1}Y$ in the laboratory system (see \cite{RD89} for the procedures for the analogous case of $(e,e^{\prime }N)$ reactions).
Using standard Feynman rules we get for the cross section:
\begin{eqnarray} \nonumber
d\sigma_{\chi} = \frac{G^2\:\cos^2 \theta_c}{2(2\pi)^5}
\:\frac{m_N \:W_{A-1} \:v_0}{k \:\varepsilon^{\prime} \:E_N \:E_{A-1}}
\:\mathcal{F}^2_{\chi} \:d^3\mathbf{k^{\prime}} \:d^3\mathbf{p_N} \:d^3\mathbf{p_{A-1}} \:\delta^4(K + P_A - K^{\prime} - P_{A-1} - P_N) \label{cross_section_general} \\
\end{eqnarray}
This form is exact in the cases where the $A-1$ system is in a bound ground state or a long-lived excited state. In other cases this form assumes that the wave function of the $A-1$ system can be factorized into center-of-mass and relative wave functions, which is not in general true for relativistic wave functions. However, since the momenta available to the $A-1$ system will generally be of the order of the Fermi momentum and the masses of the undetected fragments will tend to be large, the nuclear system will generally be treated non-relativistically and the factorization of the wave function will then be exact.
Upon integration over the unobserved residual daughter nucleus momentum $\bf{p_{A-1}}$ and energy $E_{A-1}$ one gets
\begin{equation}\label{cross_section}
\frac{d\sigma_{\chi}}{dk^{\prime }\:d\Omega _{k^{\prime }}\:d\Omega _{p_N}} =
\frac{G^2\:\cos^2 \theta_c}{2(2\pi)^5}
\:\frac{m_N \:W_{A-1}}{M^0_A}
\:\frac{p_N \:k^{\prime 2} \:v_0}{k \:\varepsilon^{\prime} \:F_{rec}}  \:\mathcal{F}^2_{\chi},
\end{equation}
where $W_{A-1}$ is defined so that $f \equiv 0$, with
\begin{equation}
f = \varepsilon + M^0_A - \varepsilon^{\prime} - \left( p^2_N + m^2_N \right)^{1/2} -\left( q^2 + p^2_N - 2\:q\:p_N\:\cos\theta_N + W^2_{A-1} \right)^{1/2}
\end{equation}
This equation is a rewriting of the energy conservation condition stated in Eq. (\ref{ener_cons}). From the function $f$ one obtains also the recoil factor $F_{rec}$ as
\begin{equation}
F_{rec} = \frac{E_N \:E_{A-1}}{M^0_A \:p_N} \:\left| \frac{\partial f}{\partial p_N} \right| = \left| 1 + \frac{\omega\:p_N - q\:E_N\:\cos\theta_N}{M^0_A\:p_N} \right|.
\end{equation}
When ERL applies, the cross section in Eq.~(\ref{cross_section}) becomes
\begin{equation}\label{cross_section_erl}
\frac{d\sigma_{\chi \:\text{[ERL]}}}{d\varepsilon^{\prime }\:d\Omega _{k^{\prime }}\:d\Omega _{p_N}} =
\frac{G^2\:\cos^2 \theta_c}{16\pi^5}
\:\frac{m_N \:W_{A-1}}{M^0_A}
\:\frac{p_N \:\varepsilon^{\prime 2} \:\cos^2(\theta/2)}{F_{rec}}  \:\mathcal{F}^2_{\chi}.
\end{equation}

\section{Conclusions \label{conclusions}}

In this study we have presented the general formalism for
{\em semi-inclusive} charged-current neutrino-nucleus reactions, {\it i.e.,} those
processes where neutrinos (antineutrinos) interact with a nuclear
target and in addition to the final-state lepton (anti-lepton) one
assumes that some other particle is also detected in coincidence.
Such processes are called semi-inclusive reactions to contrast
them from inclusive reactions where only the final-state
lepton is detected. The features summarized below highlight the generality of this formalism. We note the following:

\begin{itemize}

\item The masses of the incoming and outgoing leptons are kept, {\it
viz.,} no extreme relativistic limit has been invoked. Although for
typical kinematical situations the impact is limited when
considering scattering of active neutrinos with production of
electrons or muons, it becomes relevant for tau production, and it
can also be easily extended to study massive sterile neutrino
interactions with nuclei.

\item The scattering of both neutrinos and antineutrinos is
considered, differing just in the sign of the antisymmetric tensor
contraction contribution to the matrix element of the process.

\item No assumptions are made on the hadronic target, on the
particle emitted and detected in coincidence or on the state of the
residual, undetected hadronic system after the emission. In
particular, the latter can be in an excited bound state or be
partially or totally unbound, as long as charge and baryon numbers
are conserved.

\item The detailed
characterization of the semi-inclusive neutrino cross section is
organized in a form that makes it easy to understand as a
straightforward generalization of the well-known formalism for
inclusive \cite{DR86} and semi-inclusive \cite{RD89} electron scattering
cross sections, as well as for inclusive neutrino reactions \cite{Ama05}.
Indeed, the purely-vector semi-inclusive neutrino
responses are the same as the corresponding isovector electron
scattering responses, {\it viz.}, because of CVC. Two forms are given for the general response
structure of the cross sections, one in terms of charge-like,
longitudinal and transverse projections of the electroweak current
(the $W$s of Sects. \ref{general_tensors} and \ref{contraction})
and another in terms of invariant structure functions (the $X$s, $Y$s and $Z$s of Sect. \ref{general_tensors}).

\item Using the basic symmetries in the problem (angular momentum,
parity and four-momentum conservation) we have isolated the general
dependences on the azimuthal angle $\phi$. For instance, even
without detailed modeling, one can see how specific interference
terms in the response change sign when going from $\phi =0$ to $\phi
= \pi$. One should be clear that such interference contributions are
intrinsic to the basic semi-inclusive electroweak reaction and must
be modeled. They are not, for instance, present for inclusive
reactions, and indeed, the modeling typically used in studies of the
latter are often quite inadequate when studying semi-inclusive
scattering.

\item The general semi-inclusive response is organized into
symmetric and anti-symmetric contributions, and contributions that
are purely vector (VV), purely axial-vector (AA) and VA
interferences. For such processes, of the 16 possible response functions, the 6 underlined contributions (see Eqs.~(\ref{eqcon30}, \ref{eqcon31})) do not enter for CC$\nu$ reactions, leaving 10 distinct contributions to the semi-inclusive cross section. These in turn are built from the 17 invariant structure functions introduced in Eqs.~(\ref{eqr16}, \ref{eqr17}) (note that the term containing $Y_1$ does not contribute for CC$\nu$ reactions). In contrast, there are only 5 distinct contributions to the inclusive cross section.

\item Furthermore, the semi-inclusive responses are all functions of 4 kinematic variables, whereas the inclusive ones depend on only 2 kinematic variables. Of course, complete integrations over two of the variables in the former yield either zero for some of the interference responses or yield their inclusive counterparts.

\item Ultimately, when specific models are considered and when the neutrino fluxes commonly employed when comparing with experiment are taken into consideration, it will be necessary to integrate over the neutrino energies involved with the fluxes as weighting factors. Note, however, that this does not at all mean that one reverts to the inclusive responses. In fact, those integrations can be cast as line integrals in the $({\cal E},p)$-plane, which are not simply related to the complete integrations in that plane that would yield the inclusive responses. Indeed, such integrations leave averaged responses that depend on 3 kinematic variables and the interference responses do not integrate to zero.

\item Accordingly, the demands being placed on modeling the coincidence reactions are much greater. Where crude models such as the relativistic Fermi gas model may be acceptable for studies of inclusive scattering (to the extent that errors of perhaps 30\% are viewed as acceptable), for semi-inclusive studies many of the models being employed are certainly inapplicable, since they are incapable of predicting even roughly the correct $({\cal E},p)$-dependence of the cross section.

\end{itemize}

Furthermore, neutral-current neutrino weak interactions can also be
described by the formalism in this work upon integration over the
outgoing neutrino variables. This inclusive u-channel results in
non-vanishing responses in general,  in contrast to inclusive
t-channel reactions where integration over the momentum of the
ejected particle (of course, consistent with four-momentum
conservation) causes the responses dependent on the angle $\phi$ to
vanish (see the discussion in \cite{Ama06}).

As stated above, the formalism has been kept entirely general and any
type of coincidence reaction can be represented in terms of the
response functions introduced in this work. However, to make the
formalism clearer, we have focused on the case where the particle
detected in coincidence with the final-state muon is a nucleon. In
fact, in practical situations this is likely to be a proton so that
the semi-inclusive reactions of interest will typically be of the
type $^{A}_{Z}X(\nu_{\ell},\ell^-p)^{A-1}_{\quad Z}Y$ and
$^{A}_{Z}X(\bar{\nu}_{\ell},\ell^+p)^{A-1}_{Z-2}Y$.
A general differential cross section is given, from which a variety of
integrations can be performed; we do so over the residual daughter
nucleus variables, assuming that the incoming neutrino energy is
known, to produce a differential cross section suitable for Monte
Carlo generators. In practical situations, however, the energy of
the incoming neutrinos lies within a rather wide range, connecting
to a variety of possible dynamic regimes in the nuclear target. This
is the reason why we introduce in this work the excitation energy
and the momentum of the residual system as hadronic kinematic
variables. For given (measured) conditions such as the final lepton
and emitted nucleon momenta (both magnitude and direction, or
angles), a range of incoming neutrino energies translates into a
curve in the $({\cal E},p)$-plane that reveals which nuclear
dynamics are most relevant for the process, as for instance
multi-nucleon versus one-nucleon emission. Some care has been taken
in providing the inter-connections between the ``experimental"
kinematic variables (energies and momenta of the detected particles)
and the ``nuclear" kinematic variables, $p$ and ${\cal E}$, since
the response of the nucleus is a rapidly-varying function of the
latter.

Our plan for work already in progress is to study specific reactions
involving particular nuclei. In doing so it is clearly essential to
understand where the dominant regions in the $({\cal E},p)$-plane
lie to be able to predict the semi-inclusive (and also inclusive)
neutrino cross sections with sufficient confidence.

\begin{acknowledgments}
This research was supported by a Marie Curie International Outgoing Fellowship within the 7th European Community Framework Programme and by MINECO (Spain) under Research Grant No. FIS2011Ð23565 (O. Moreno). Also supported in part by the US Department of Energy under cooperative agreement DE-FC02-94ER40818 (T. W. Donnelly), and by the US Department of Energy under Contract No. DE-AC05-06OR23177 and the U.S. Department of Energy cooperative research agreement DE-AC05-84ER40150 (J. W. Van Orden).
\end{acknowledgments}

\end{document}